\newcommand{\uno}{{\bf 1}}
\newcommand{\dos}{{\bf 2}}

\newcommand{\cuatro}{{\bf 4}}

\newcommand{\ocho}{{\bf 8}}
\newcommand{\diez}{{\bf 10}}

\newcommand{\treintaycinco}{{\bf 35}}
\newcommand{\cincuentayseis}{{\bf 56}}
\newcommand{\setenta}{{\bf 70}}

\newcommand{\setecientos}{{\bf 700}}
\newcommand{\milcientotreintaycuatro}{{\bf 1134}}

\documentclass[epj]{svjour}
\usepackage{graphics}
\begin{document}
\title{Resonances and the Weinberg--Tomozawa 56-baryon --35-meson
 interaction.}  \author{ C. Garc{\'\i}a-Recio\inst{1}, J. Nieves\inst{1} \and L.L. Salcedo\inst{1}% etc
}                     % Do not remove
\institute{Departamento de
 F{\'\i}sica At\'omica, Molecular y Nuclear, Universidad de Granada,
 E-18071 Granada, Spain}
\date{Received: date / Revised version: date}
\abstract{Vector meson degrees of freedom are incorporated into the
  Weinberg-Tomozawa (WT) meson-baryon chiral Lagrangian by using a
  scheme which relies on spin--flavor SU(6) symmetry. The
  corresponding Bethe-Salpeter approximation successfully reproduces
  previous SU(3)--flavor WT results for the lowest-lying s--wave
  negative parity baryon resonances, and it also provides 
  some information on the dynamics of the heavier ones. Moreover, it
  also predicts the existence of an isoscalar spin-parity $\frac32^-$
  $K^*N$ bound state (strangeness +1) with a mass around 1.7--1.8 GeV,
  unstable through $K^*$ decay.  Neglecting d-wave KN decays, this
  state turns out to be quite narrow ($\Gamma \le 15$ MeV) and it
  might provide clear signals in reactions like $\gamma p \to \bar K^0
  p K^+\pi^-$ by looking at the three body $p K^+\pi^-$ invariant
  mass.
 % end of PACS codes
} %end of abstract
\maketitle
\section{Introduction}
\label{intro}
We present results obtained from a scheme where it is assumed that the
light quark--light quark interaction is approximately spin independent
as well as SU(3) independent. This corresponds to treating the six
states of a light quark ($u$, $d$ or $s$ with spin up, $\uparrow$, or
down, $\downarrow$) as equivalent, and leads us to the invariance
group SU(6). Despite the fact that the no--go Coleman--Mandula
theorem~\cite{Coleman:1967ad} forbids this hybrid symmetry (mixing the
compact, purely internal flavor symmetry, with the noncompact Poincare
symmetry of spin angular momentum) to be exact, there exist several
SU(6) predictions (relative closeness of baryon octet and decuplet
masses, the axial current coefficient ratio $F/D=2/3$, the magnetic
moment ratio $\mu_p/\mu_n=-3/2$) which are remarkably well satisfied
in nature~\cite{Lebed:1994ga}. This suggests that SU(6) could be a
good approximate symmetry. Though in general the spin--flavor symmetry
is not exact for excited baryons even in the large $N_c$ limit (being
$N_c$ the number of colors)\footnote{In the large $N_c$
  limit~\cite{'tHooft:1973jz,Witten:1979kh}, there exists an exact
  spin--flavor symmetry for ground state baryons.}, in the real world
($N_c=3$), the zeroth order spin--flavor symmetry breaking turns out
to be similar in magnitude to ${\cal O} (N_c^{-1})$ breaking
effects~\cite{Goity:2002pu}. Spin-flavor symmetry in the meson sector
is not a direct consequence of large $N_c$ QCD either.  However vector
mesons ($K^*, \rho,\omega, {\bar K}^{*}, \phi$) do exist and they are
known to play a relevant role in hadronic physics.  Inescapably, they
will couple to baryons and will presumably influence the properties of
the baryonic resonances.  Lacking better theoretically founded models
to include vector mesons, we regard the spin-flavor symmetric scenario
as reasonable first step. The large $N_c$ consequences of this scheme
have been pursued in \cite{Garcia-Recio:2006wb}.

We  will consider the $s$-wave interaction between the SU(6)
lowest--lying meson multiplet ($\treintaycinco$) and the lowest--lying
baryons ($\cincuentayseis$-plet) at low energies. The meson multiplet
contains the octet of pseudoscalar ($K, \pi,\eta, {\bar K}$) and the
nonet of vector ($K^*, \rho,\omega, {\bar K}^{*}, \phi$) mesons, while
the baryon one is constructed from the $(N,\Sigma,\Lambda, \Xi)$ octet
of spin--$1/2$ baryons and the ($\Delta$, $\Sigma^*$, $\Xi^*$,
$\Omega$) decuplet of spin--$3/2$ baryons. Assuming that the
$s$-wave effective meson--baryon Hamiltonian is SU(6) invariant, and
since the SU(6) decomposition of the product of the $\treintaycinco$
(meson) and $\cincuentayseis$ (baryon) representations yields
\begin{eqnarray}
\treintaycinco \otimes \cincuentayseis = \cincuentayseis \oplus 
\setenta \oplus \setecientos \oplus \milcientotreintaycuatro,
\label{eq:1}
\end{eqnarray}
there are only four, Wigner-Eckart irreducible matrix elements
(WEIME's), free functions of the meson--baryon Mandelstam variable
$s$. Similar ideas were already explored in the late sixties, within
the effective range approximation~\cite{Carey:1968}.  In this work,
based on the findings of Ref.~\cite{Garcia-Recio:2005hy}, two major
improvements have been introduced: i) The use of the underlying Chiral
Symmetry (CS), which would allow to determine the value of the SU(6)
irreducible matrix elements from the Weinberg-Tomozawa (WT)
interaction (leading term of the chiral Lagrangian involving Goldstone
bosons and the octet of spin--$1/2$ baryons), ii) the use of Bethe-Salpeter
Equations (BSE)~\cite{Nieves:1998hp,Nieves:1999bx}, in coupled channels
and with an appropriated Renormalization Scheme
(RS)~\cite{Lutz:2003fm,Garcia-Recio:2003ks}, to determine the
scattering amplitudes, going thus beyond the effective range
approximation.

\section{SU(6) extension of the meson--baryon WT interaction }
\label{sec:1}
We will work with well defined total isospin ($I$), angular momentum
($J$) and hypercharge ($Y$) meson--baryon states constructed out of
the SU(6) $\treintaycinco$ (mesons) and $\cincuentayseis$ (baryon)
multiplets.  By imposing that the effective $s$-wave meson--baryon
{\it potential} ($V$) is a SU(6) invariant operator, we
find~\cite{Garcia-Recio:2005hy}
\begin{eqnarray}
\langle {\cal M}^\prime  {\cal B}^\prime ; JIY | V
| {\cal M} {\cal B} ; JIY \rangle &=&
\sum_{\phi} V_{\phi}(s) {\cal P}_{{\cal M}{\cal B}, {\cal M}^\prime 
{\cal B}^\prime }^{\phi,JIY} ,
\label{eq:su6}
\end{eqnarray}
where ${\cal M}\equiv \left [(\mu_M)_{2J_M+1}, I_M, Y_M\right]$ stands
for meson states and similarly ${\cal B}$ for baryon ones and the
labels $\mu$ and $\phi$ denote SU(3) and SU(6) representations,
respectively. In the above equation $\phi$ runs over the
$\cincuentayseis$, $\setenta$, $\setecientos$ and
$\milcientotreintaycuatro$ irreducible representations (irreps), as
inferred from Eq.~(\ref{eq:1}). The projectors are given in terms of
SU(3) isoscalar \cite{deSwart:1963gc}, and the SU(6)--multiplet
coupling~\cite{Carter:1969,Carter:1969erratum} factors,
\begin{eqnarray}
&&
{\cal P}_{{\cal M}{\cal B}, {\cal M}^\prime 
{\cal B}^\prime }^{\phi,JIY}
= \sum_{\mu,\alpha} 
\left( \begin{array}{cc|c} \treintaycinco& \cincuentayseis&
   \phi \\ \mu_M J_M & \mu_B J_B &
   \mu J \alpha \end{array}\right)\nonumber\\
&\times& 
\left( \begin{array}{cc|c}
 \mu_M& \mu_B& \mu \\ 
  I_M Y_M & I_B Y_B & I Y\end{array}\right)
\left( \begin{array}{cc|c} \mu^\prime_{M^\prime}& \mu_{B^\prime}^\prime&
   \mu \\ I^\prime_{M^\prime}Y^\prime_{M^\prime} &
 I^\prime_{B^\prime} Y^\prime_{B^\prime} &
   IY\end{array}\right) 
\nonumber\\
&\times&
\left( \begin{array}{cc|c} \treintaycinco& \cincuentayseis&
   \phi \\ \mu'_{M'} J'_{M'} & \mu'_{B'} J'_{B'} &
   \mu J \alpha \end{array}\right).
 \phantom{hhhhhhhhhhhhh}
\end{eqnarray}
where $\alpha$ accounts for the multiplicity of each of the
$\mu_{2J+1}$ SU(3) multiplets of spin $J$ (for $L=0$, $J$ is given by
the total spin of the meson--baryon system) entering in the
representation $\phi$. The SU(6) WEIME's, $V_{\phi}(s)$, might be
constrained by demanding that the above interaction, when restricted to
the Goldstone pseudoscalar meson and the lowest $J^P=\frac12^+$ baryon
octet space, will reduce to that deduced from SU(3) chiral
symmetry. At leading order in the chiral expansion, this latter one is
obtained from the WT Lagrangian, which besides hadron masses only
depends on the $f\simeq 93\,$MeV the pion weak decay constant, and it
can be exactly recovered from Eq.~(\ref{eq:su6}) by setting the WEIME's as
follows~\cite{Garcia-Recio:2005hy} 
\begin{equation}
V_{\phi}(s) = \bar\lambda_{\phi}
\frac{\sqrt{s}-M}{2\,f^2}\,,
\label{eq:vsu6} 
\end{equation}
with $\bar\lambda_{\cincuentayseis}=-12$,
$\bar\lambda_{\setenta}=-18$, $\bar\lambda_{\setecientos}=6$ and
$\bar\lambda_{\milcientotreintaycuatro}=-2$ and $M$ the common octet
and decuplet baryon mass\footnote{The SU(6) extension thus obtained
(Eqs.~(\ref{eq:su6}) and~(\ref{eq:vsu6})) also leads to the {\it
potentials} used in Ref.~\cite{Kolomeitsev:2003kt,Sarkar:2004jh} to
study the
($\protect\diez_\protect\cuatro$)baryon--($\protect\ocho_\protect\uno$)meson
sector.}. This is not a trivial fact and it is intimately linked to
the group structure of the WT term. Indeed, the underlying reason for
this is CS, since the WT Lagrangian is not just SU(3) symmetric but
also chiral (${\rm SU}_L(3)\otimes{\rm SU}_R(3)$)
invariant~\cite{Garcia-Recio:2005hy}.
\section{Meson--baryon scattering matrix}
We solve the coupled channel BSE with an interaction kernel determined
by Eqs.~(\ref{eq:su6}) and~(\ref{eq:vsu6}). In a given $JIY$
sector, the solution for the coupled channel $s$-wave scattering
amplitude, $T^{J}_{IY}(\sqrt{s})$ (normalized as the $t$ matrix
defined in Eq.~(33) of \cite{Nieves:2001wt}), in the {\it on-shell}
scheme~\cite{Nieves:1998hp,Nieves:1999bx,%
Oset:1997it,Lutz:2003fm,Jido:2003cb}
reads,
\begin{eqnarray}
T^J_{IY}(\sqrt{s}) &=& \frac{1}{1-
V^J_{IY}(\sqrt{s})\,J^J_{IY}(\sqrt{s})}\,V^J_{IY}(\sqrt{s})
 \label{eq:scat-eq}
\end{eqnarray}
with $V^J_{IY}(\sqrt{s}) = \langle {\cal M}^\prime  {\cal B}^\prime ; JIY
| V| {\cal M} {\cal B} ; JIY \rangle
\,,$ and $J^J_{IY}(\sqrt{s})$ a diagonal matrix of loop
functions~\protect\cite{Nieves:2001wt,Garcia-Recio:2002td}. Those are
logarithmically divergent and hence to make them finite an ultraviolet
(UV) cutoff or a subtraction point $\mu^{JIY}_i$, such that
\begin{equation}
 \left[J^J_{IY}(\sqrt{s}=\mu^{JIY}_i)\right]_{ii} = 0
\end{equation}
with the index $i$ running in the coupled channel space, is needed.

\begin{figure}
\resizebox{0.5\textwidth}{!}{%
  \includegraphics{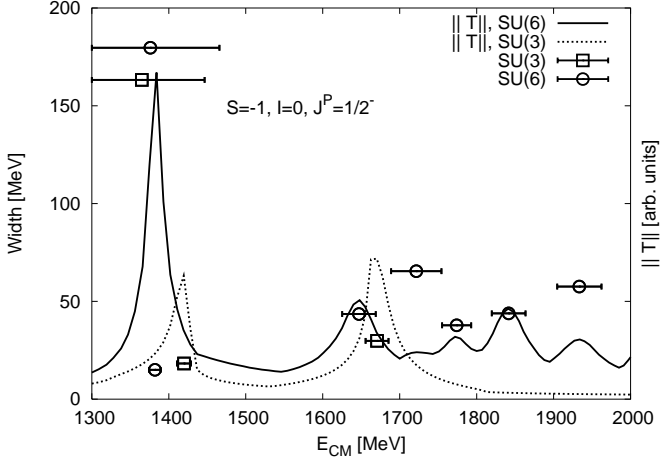}
}
%\vspace{5cm}       % Give the correct figure height in cm
\caption{Spin--parity $J^P=\frac12^-$ resonance properties in the 
$I=0, S=-1\ (Y=0)$ sector. Solid and dotted lines stand for the SU(6) and SU(3)
$T-$matrix norms, respectively. The norm is calculated as $||T||={\rm
Max} \left (\sum_{j=1}^n |T_{ij}|, i=1,\ldots n\right )$ where $i,j$
run over the number of coupled channels. Finally the points with
error-bars are defined from the masses and widths of the found
resonances as $(M_R\pm\Gamma_R/2,
\Gamma_R)$.}
\label{fig:1}       % Give a unique label
\end{figure}

By setting $\mu^{JIY}_i=\sqrt{M^2_i+m^2_i}$, with $M_i$ and $m_i$ the
masses of the baryon and meson entering in the channel $i$ of the sector $JIY$,
we recover previous results, deduced from the WT SU(3) chiral
Lagrangian~\cite{Lutz:2003fm,Garcia-Recio:2003ks,Nieves:2001wt,Oset:1997it,Jido:2003cb,Garcia-Recio:2002td},
and make new predictions. For instance in the $I=0, S=-1\ (Y=0)$ sector,
looking for poles in the second Riemann sheet,
\begin{itemize}
 \item For $J^P=\frac12^-$, we obtain the $\Lambda(1390),
\Lambda(1405), \Lambda(1670)$ resonances,  but we also find strength around
1800 MeV, which might correspond to the three star $\Lambda(1800)$,
which has a sizeable $N{\bar K}^*$ coupling (see Fig.~\ref{fig:1}).

\item For $J^P=\frac32^-$, we get signals for
some d-wave resonances (not accessible with the SU(3) WT chiral
  Lagrangian): the four star  $\Lambda(1520)$ and
  $\Lambda(1690)$ states with large couplings to the $\Sigma^* \pi$ channel,
  and the $\Lambda(2325)$ resonance which couples to $\Lambda \omega$
  channel.
\end{itemize}
Similar results are found for the other $I,Y$ sectors. SU(6) symmetry
breaking effects such that the use of a common weak decay constant ($f$)
for all  channels lead to changes of the order of 30\% in resonance
widths and excitation energies. These uncertainties are similar to those
stemming from the use of different UV cutoffs or subtraction points to
renormalize the amplitudes.
\section{Exotic states}

The SU(6) meson--baryon interaction constructed in Sect.~\ref{sec:1}
turns out to be attractive in the $\milcientotreintaycuatro-$irrep space
($\bar\lambda_{\milcientotreintaycuatro}=-2$), which might lead to the
existence of some exotic states. For instance,  in the $Y=-3,
I=J=1/2$ sector, we have {\bf $|{\bar K}^* \Omega \rangle=
|\milcientotreintaycuatro; \treintaycinco_\dos \rangle $} or for
$Y=+2,I=0,J=3/2$ {\bf $|K^*N\rangle= - |\milcientotreintaycuatro;
\diez^*_\cuatro \rangle $}, hence one finds attractive ${\bar K}^* \Omega$
or $K^*N$ interactions with these exotic quantum numbers. Whether these
interactions are strong enough or not to bind the meson--baryon system
will depend on the RS employed to make finite the BSE and on the
nature of the further $s-$ and $d-$wave contributions to the
interaction matrix. 

Neglecting any further correction to the {\it potential}, we present
here results for the $K^*N$ system in the $I=0,J=3/2$ sector (see
Ref.~\cite{Garcia-Recio:2005hy} for some more details). We find a pole
in the first Riemann sheet corresponding to a $K^* N$ bound state
which we call $\Theta^{*+}$. This state is unstable since the $K^*$
decays into $K\pi$.  to estimate the $\Theta^{*+}$ width, we model the
$\Theta^{*}N K^*$ coupling as
\begin{eqnarray}
{\cal L}_{\Theta^* N K^*}&=& 
-\frac{g}{\sqrt{2}} 
\overline{\Theta}^\mu 
 \left(  K^{*0}_\mu p - K^{*+}_\mu n \right) 
 + {\rm h.c.},
\end{eqnarray}
where $\Theta^\mu$ is a Rarita-Schwinger field, $p$ and $n$ the
nucleon fields, $ K^{*0}_\mu$ and $K^{*+}_\mu $ the Proca fields which
annhilate and create neutral and charged $K^*$ and ${\bar K}^*$
mesons. The subsequent $K^*$ decay is described following
Ref.~\cite{Ecker:1989yg}. The coupling $g$ is determined by the residue
at the pole of $T^{\frac32}_{02}$ [i.e., $T^{\frac32}_{02} \approx g^2
\times 2M_{\Theta^*}/(s-M^2_{\Theta^*})$].

\begin{figure}
\begin{center}
\makebox[0pt]{\input{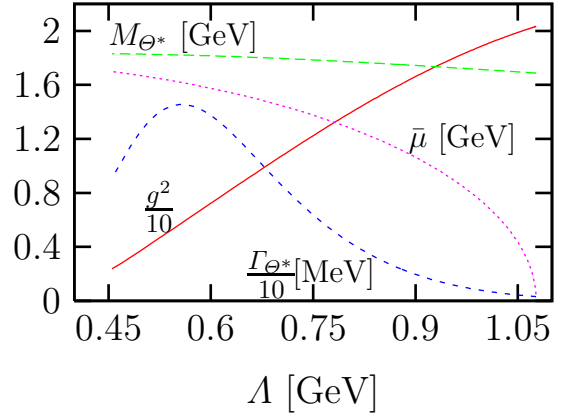}}
\end{center}
%\vspace{-0.5cm}
\caption{Resonance $\Theta^{*+}$ properties as a function of the
  UV cutoff $\Lambda $ or the subtraction scale ${\bar \mu}$.}
\label{fig:fig1}
\end{figure}

Resonance mass, residue and width depend on the RS employed. We have
used an UV cutoff ($\Lambda$) to evaluate the loop function
$J(\sqrt{s})$, which is equivalent to choose an scale ${\bar \mu}$
such that $J(\sqrt{s}={\bar \mu})=0$. Results are shown in
Fig.~\ref{fig:fig1}. For ${\bar \mu}$ ranging from $0.05\,$GeV
($\Lambda\approx 1.08\,$GeV) to $1.7\,$GeV ($\Lambda\approx
0.46\,$GeV) the resonance mass (width) varies from $1.688\,$GeV
($0.3\,$MeV), close to the $(M_N+m_\pi+m_K)$ threshold, to
$1.831\,$GeV ($9\,$MeV, but the width does not grow monotonously, see
figure), close to the $(M_N+m_{K^*})$ threshold.
Other mechanisms for $K^*N$ scattering ($d$-wave ${K} N$, ${K}^* N$
contributions, $u$-channel pole graph,  single pion exchange between
$K^*$ and $N$, sequential exchange of two pions with an intermediate
$K$ meson, corresponding to a box graph $K^*N\to KN \to K^*N$,  \ldots) might
quantitatively modify these results. However, we do not expect such
corrections to be large enough to affect the existence of the
$\Theta^*$ pentaquark~\cite{Garcia-Recio:2005hy}.  Possible production
and identification mechanisms for this resonance could be found in
reactions like $\gamma p \to {\bar K}^0 p K^+ \pi^-$ by measuring the
three body $p K^+ \pi^-$ invariant mass.

\begin{acknowledgement}
This work was supported by DGI, FEDER, UE and Junta de Andaluc{\'\i}a funds 
(FIS2005-00810, HPRN-CT-2002-00311, FQM225).
\end{acknowledgement}

%
% BibTeX users please use
% \bibliographystyle{}
% \bibliography{}
		
%

\end{document}